# Phase Transition and Electronic Structure Investigation of $MoS_2$-rGO Nanocomposite Decorated with AuNPs.

**Yunier Garcia-Basabe[1,*], Gabriela F. Peixoto[2], Daniel Grasseschi[2,3], Eric C. Romani[4], Flávio C. Vicentin[5], Cesar E. P. Villegas[6], Alexandre. R. Rocha[7] and Dunieskys G. Larrude[3]**

[1] *Universidade Federal da Integração Latino-Americana, UNILA, 85867-970, Foz do Iguaçu, Brazil.*
[2] *Inorganic Chemistry Department, Chemistry Institute, Federal University of Rio de Janeiro (UFRJ), 21941-909, Rio de Janeiro, Brazil*
[3] *MackGraphe-Graphene and Nanomaterial Research Center, Mackenzie Presbyterian University, 01302-907, São Paulo, Brasil.*
[4] *SENAI Innovation Institute for Virtual Production Systems, 20911-210, Rio de Janeiro, Brazil.*
[5] *Brazilian Synchrotron Light Laboratory (LNLS), Brazilian Center for Research in Energy and Materials (CNPEM), 13083-970, Campinas, Sao Paulo, Brazil.*
[6] *Departamento de Ciencias, Universidad Privada del Norte, Av. Andrés Belaunde Cdra. 10 s/n, 15324, Comas, Lima, Peru.*
[7] *Instituto de Física Teórica, State University of São Paulo (UNESP), 01049-010, São Paulo, Brazil.*

*Corresponding Author:

Prof. Dr. Yunier Garcia-Basabe

E-mail address: yunier.basabe@unila.edu.br and yunierbasabe26@gmail.com

ORCID iD: 0000-0001-5683-0108

ABSTRACT

In this work a simple approach to transform $MoS_2$ from its metallic (1T' to semiconductor 2H) character via gold nanoparticle surface decoration of a $MoS_2$ graphene oxide (rGO) nanocomposite is proposed. The possible mechanism to this phase transformation was investigated using different spectroscopy techniques, and supported by density functional theory theoretical calculations. A mixture of the 1T'- and 2H-$MoS_2$ phases was observed from the Raman and Mo 3d High Resolution X-ray photoelectron (HRXPS) spectra analysis in the $MoS_2$-rGO nanocomposite. After surface decoration with gold nanoparticles the concentration of the 1T' phase decreases making evident a phase transformation. According to Raman and valence band spectra analyses, the AuNPs induces a p-type doping in $MoS_2$-rGO nanocomposite. We proposed as a main mechanism to the $MoS_2$ phase transformation the electron transfer from Mo $4d_{xy,xz,yz}$ in 1T' phase to AuNPs conduction band. At the same time, the unoccupied electronic structure was investigated from S $K$-edge Near Edge X-Ray Absorption Fine Structure (NEXAFS) spectroscopy. Finally, the electronic coupling between unoccupied electronic states was investigated by the core hole clock approach using Resonant Auger spectroscopy (RAS), showing that AuNPs affect mainly the $MoS_2$ electronic states close to Fermi level.

# 1 INTRODUCTION

Over the past few years, two-dimensional transition metal dichalcogenides (TMDs) have become attractive options for electronic devices, transistors, biosensors, as well as for applications in catalysis and energy storage devices [1-3]. Among the TMD family $MoS_2$ is the most investigated due to its unique physical and chemical properties [4,5]. A singular feature of $MoS_2$ layered materials is the structural polymorphism, which modulates its electronic structure. In particular, the 2H and 1T phases with different arrangement of the S and Mo atoms have been reported for $MoS_2$ crystals [6]. The 2H-$MoS_2$ is the most stable structure, where each layer is constructed from the Mo atom sandwiched between two S atoms in a trigonal prismatic ($D_{3h}$) geometry. On the other hand, the 1T-$MoS_2$ phase is based on Mo atoms octahedrally ($O_h$) coordinated to S atoms. $MoS_2$ in 1T phase is metastable (easily converted to the stable 2H phase) while in the octahedral distorted configuration (1T') it turns into a phasethat is thermodynamically stable [7]. The trigonal prismatic symmetry in 2H-$MoS_2$ induces a semiconductor electronic structure with monolayer direct band gap of ~1.8 eV, [1,8] while the octahedral coordination of the Mo atoms in the 1T'-$MoS_2$ phase gives rise to a metallic behavior with good conductivity [1,9]. Recently, several works about the physical and chemical treatment of $MoS_2$ to obtain a reversible transition between these two phases have been reported [6,10-12]. In most of them, the 1T'-$MoS_2$ phase presents superior electrocatalytic properties, while 2H-$MoS_2$ is an excellent candidate for optoelectronic applications [13-15].

While isolated $MoS_2$ has attractive properties, the formation of $MoS_2$-graphene nanocomposites could be an excellent approach to improve its performance for different device applications [16-20] as the mechanical flexibility, electronic conductivity and chemical stability of graphene could added to the properties of isolated $MoS_2$. Recently it was reported that p-$MoS_2$/n-rGO nanocomposites have better photocatalytic activity in hydrogen generation when compared to pristine $MoS_2$ [21,22]. At the same time noble metal nanoparticles (NPs) decorating the $MoS_2$-graphene nanocomposite surface could potentially extend its functionalities as novel catalytic, magnetic, and optoelectronic nanomaterials even further [2,23-26]. Among these nanoparticles, Gold (AuNPs) is of special interest due to their non-toxicity and excellent stability. In fact, Au-$MoS_2$ composite synthesis, electrical and thermal properties and its application on photocatalytic water splitting and hydrogen production has been the subject of a number of works [27-31].

Nevertheless, the knowledge of the electronic structure of the $MoS_2$-graphene nanocomposites is a crucial piece of information for better understanding their biosensing, photonics, catalytic and optoelectronic device performance. In particular, controlling the different phases with metallic and semiconductor behavior of the nanocomposite is of utmost importance for appropriately tailoring properties, and subsequent applications. In this sense, in the present work, the effect of AuNPs decoration on the stability of $MoS_2$ (1T' and 2H) phases and in the electronic structure of a $MoS_2$-rGO nanocomposite was investigated. The local electronic structure of either the pristine nanocomposite, and decorated with AuNPs was studied using High Resolution X-ray Photoelectron Spectroscopy (HRXPS), Near Edge X-Ray Absorption Fine Structure (NEXAFS), Resonant Auger Spectroscopy (RAS), and Density Functional Theory simulations (DFT). The non-local electronic structure was investigated by Valence band and Raman spectra. A mixture of the 1T' and 2H phases of $MoS_2$ was observed from the Raman spectrum in the solvothermal-synthesized $MoS_2$-rGO nanocomposite and corroborated from Mo 3d HRXPS spectra analysis. A decrease of the Mo$^{4+}$ (1T') species is shown after surface decoration with AuNPs, making evident a phase transformation from 1T' to 2H-MoS$_2$. We identify surface electron transfer processes from MoS$_2$-rGO to AuNPs as the main mechanism for this phase transformation using Raman and XPS valence band spectra analysis. At the same time, the unoccupied electronic structure was investigated from S K-edge NEXAFS spectroscopy. The degree of delocalization and/or electronic coupling between unoccupied electronic states was investigated by the core hole clock approach showing that AuNPs affect the electronic states close to the Fermi level.

## 2. EXPERIMENTAL DETAILS

*2.1 Synthesis procedure*

Graphene oxide synthesis was carried out by a modified Hummers method in a calorimetric automated reactor. 0.75 g of pristine graphite flakes (Aldrich) was added to 90 mL of sulfuric acid (Sigma-Aldrich) e 10 ml of phosphoric acid followed by the slow addition of 4.5 g of potassium permanganate. This system was kept under stirring for two hours at 50°C. Subsequently, the temperature was decrease for 0°C then poured on 400ml of ice, followed by 2 mL of hydrogen peroxide (30 vol% on water, Sigma-Aldrich). The solid material was purified by centrifugation and washed with different amounts of water, hydrochloric acid solution (10 vol%), ethanol, acetone and DI water again. The black and solid material (Graphite oxide – Gr-O) was dried under vacuum overnight. An aqueous suspension of graphite oxide was then sonicated (50W, 40 kHZ) in DI water for 2 min. The resulting brown and stable suspension was found to be Graphene Oxide.

For the preparation of $MoS_2$-rGO nanocomposite, 25 mg of $(NH_4)_2MoS_4$ was added to 10 mg of graphene oxide (GO) dispersed in 10 ml of dimethylformamide (DMF). The mixture was sonicated at room temperature for approximately 10 min until a clear and homogeneous solution was achieved. After that, 0.1 ml of $N_2H_4 \cdot H_2O$ was added. The reaction solution was further sonicated for 30 min before transferred to a 150 mL Teflon-lined autoclave. The autoclave was sealed and heated in an oven at 200 °C for 10 h. The product was collected by centrifugation at 8000 rpm for 5 min and resuspended in DI water. This washing process was repeated 5 times to remove the DMF.

Gold nanoparticles were synthesized according the procedure described by Grasseschi et al [32]. All glassware were previously cleaned with aqua regia (3HCl:1HNO$_3$), in order to prevent any possible influence of contaminants in gold reduction process. The synthesis was performed, by heating 50 mL of gold solution HAuCl$_4$ solution (5.08 x 10$^{-4}$ mol L$^{-1}$) in a 125 mL round bottom flask, and refluxing, using a computer controlled system, for keeping the temperature (± 1°C) and stirring rates constants. Then, 1.7 mL of aqueous sodium citrate solution (0.038 mol L$^{-1}$) was added dropwise (100 µLs$^{-1}$), while the mechanical stirring and temperature were kept constant, at 1150 rpm and 95°C, respectively. After 5 min, the solution was cooled to room temperature.

A simple method was utilized to obtain the MoS$_2$-rGO-AuNPs nanocomposite. Thin film of MoS$_2$-rGO was deposited on the SiO$_2$/Si substrate by the drop casting method. After that, two drops of 25µL of the AuNPs solution were deposited to cover the entire film surface. The sample was dried in air at room temperature for 24 h.

*2.2 Characterization*

In order to obtain morphological information of the samples, transmission electron microscopy (TEM, JEOL JEM 2100), Scanning transmission electron microscopy (STEM) using a high angular dark field detector (HAADF, FEI Titan 80-300) and Scanning electron microscopy (SEM, JEOL JSM-7800F and Phenom ProX) were performed. The Raman spectra were collected using a WITec Alpha 300R confocal Raman imaging microscope using a laser line centered at 532 nm and 0.508 mW of power. For reducing the sample locations effect, the Raman spectra in both samples were collected at four different points and the reproducibility of results were observed. X-ray photoelectron spectroscopy (XPS) (K-Alpha Thermo Scientific) measurements and analyses were conducted to investigate the local electronic structure of the chemical species on MoS2-rGO and MoS2-rGO-AuNPs nanocomposites. The monochromatized Al Kα (hv=1486.6 eV) excitation energy with 400 μm X-ray spot size was used for XPS measurements. The electron energy analyzer was operated at constant pass energy of 25 eV, and 200 eV for high resolution (HR) and survey spectra, respectively. An instrumental energy resolution of 0.25 eV determined from $Ag_{5/2}$ peak from pure silver was used for HRXPS experiments. A flood gun source of low energy electrons and $Ar^+$ ions was used during all measurements in order to prevent surface charging. The high-resolution XPS spectra were processed using Thermo Scientific's Avantage XPS software package (version 4.61) and using a linear combination of Gaussian and Lorentzian line shapes for the spectra fitting, with a Shirley function for background correction. The binding energy scale was calibrated by attributing an energy value of 284.5 eV to C-C/C-H species in HR C1s spectrum.

The electronic structure for unoccupied states was investigated from X-ray absorption (XAS) and Resonant Auger (RAS) spectroscopies experiments. XAS and RAS were measured at the soft X-ray spectroscopy (SXS) beamline at the Brazilian Synchrotron Light Source (LNLS) facilities [33]. A Si(111) double-crystal monochromator in high resolution condition with energy bandwidth of 0.48 eV was used to cover Sulphur K-edge. Near Edge X-Ray Absorption Fine Structure spectra were recorded in the total electron yield mode (TEY). The final data was normalized by this photon flux spectrum (estimated from Au mesh) to correct the fluctuations in beam intensity. Presented spectra are averages from at least three scans, background corrected by a linear pre-edge subtraction and linear regression beyond the edge. Multiple-scan RAS spectra were collected inside an ultrahigh vacuum chamber (UHV) with a base pressure of $10^{-9}$ mbar using a hemispherical electron energy analyzer (Specs model phoibos 150) with 45º take-off direction of Auger electrons and 25 eV of pass energy. The total energy resolution was 0.76 eV. RAS spectra fittings were performed using Pseudo-Voigt profile functions (linear combination of Gaussian (G) and Lorentzian (L) functions) and the background was corrected using a Shirley function.

*2.3 Theoretical Simulations*

Plane-wave density functional theory [34,35] was used to obtain the electronic ground-state using the Perdew-Burke-Ernzerhof (PBE) generalized gradient exchange-correlation functional, [36] implemented in the Quantum-Espresso (QE) package, [37] and Non-empirical van der Waals corrections proposed by Tkachenko and Scheffler [38] were included in all calculation. The kinetic energy cutoff was set to 90 Ry and a mesh of 14 x 14 x 1 k-points in the Monkhorst-Pack scheme was used. The structures were fully optimized to their equilibrium positions with forces smaller than 0.01 eV/Å. A vacuum region of 15 Å was used to suppress the coupling between neighboring slabs. The interfaces were modeled by choosing an Au (111) surface with 4 atomic layers. Subsequently a fully relaxed 1T'-$MoS_2$ (2H-$MoS_2$) monolayer is adsorbed on the Au(111) surface following the arrangement proposed by Zhou et al [39]. By doing this, both the Au(111) and 2H-$MoS_2$ unit cell is matched with the ideal lattice constant of 1T'-$MoS_2$ monolayers. We should mention that this five-layer model has been adopted in previous works describing well some features observed in experiments [40,41].

## 3. RESULTS AND DISCUSSION

The morphology of the $MoS_2$-rGO and $MoS_2$-rGO-AuNPs nanocomposites were characterized by SEM, TEM e STEM microscopy images and are depicted in Figure 1. . As seen in SEM (Figure 1a), TEM (Figure 1b) and STEM (Figure 1c) images, the $MoS_2$-rGO nanocomposite consist of $MoS_2$ layer covering the surface of the rGO sheets. $MoS_2$ basal planes are clearly observed from high magnification TEM image shown in Figure 1b. Moreover, the HRTEM image reveals a large number of $MoS_2$ layers with interlayer distance of 0.62 nm (corresponding to the (002) plane) stacked together. Energy dispersive X-ray spectra (EDX) (Figure 1g and Figure S1 on the Supporting Information File (SI)) shows that Mo, S, C and O elements were distributed uniformly on the entire flake. Some regions show the formation of spherical particles of $MoS_2$ with an average diameter of approximately a hundred nanometers covering the rGO surfaces (Figure S1).

The effect of AuNPs decorating the $MoS_2$-rGO surface is presented in Figure 1d, showing a substantial coverage of AuNPs (bright points) in the surface of the $MoS_2$ and promoting the formation wrinkled nanosheets. The TEM and STEM images on Figure 1e and f show that the AuNPs are spherical with an average size of 23.5 ± 3.5 nm with good size dispersion, as indicated by the size distribution on Figure 1h. The EDX spectra record at different points (Figure 1g and Figure S1 and S2) confirm the presence of gold on the $MoS_2$-rGO surface.

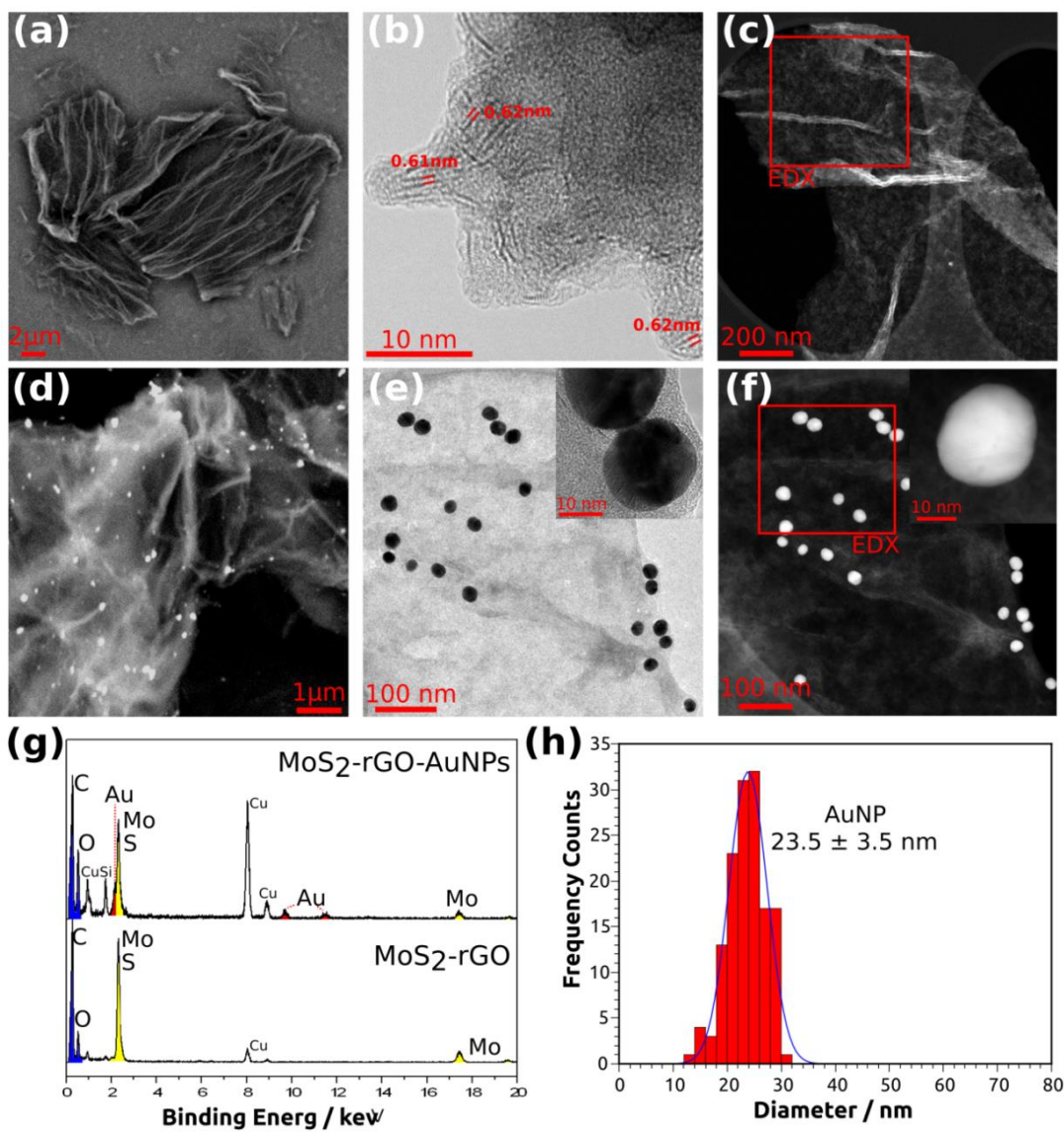

**Figure 1.** SEM images of MoS$_2$-rGO (a) and MoS$_2$-rGO-AuNPs (d) nanocomposites. TEM images of MoS$_2$-rGO (b) and MoS$_2$-rGO-AuNPs (e) nanocomposites. STEM - HAADF images of MoS$_2$-rGO (c) and MoS$_2$-rGO-AuNPs (f) nanocomposite. EDX spectra of MoS$_2$-rGO and the MoS$_2$-rGO-AuNPs nanocomposites of the selected areas on c and f (g). AuNPs size distribution (h).

Figure 2 shows the Raman spectra of MoS$_2$-rGO and MoS$_2$-rGO-AuNPs nanocomposites. These spectra exhibit the D, G and D´ peaks, vibrational modes characteristic of graphitic materials [39,42]. The G-band corresponds to the in-plane optical mode due to bond stretching of graphitic sp$^2$ carbon atoms. The D-band originates from the lattice defect and lattice distortion, while D´-band is attributed to inter and intra-valley scattering processes. The three peaks observed at 327, 379.1 and 407 cm$^{-1}$ (Figure. 2(b)) correspond to the in-plane $E_{1g}$ and $E_{2g}^1$ and out-plane $A_{1g}$ vibrational modes of the 2H-MoS$_2$, respectively [42-44]. The $A_{1g}$ peak position and the $A_{1g}$-$E_{2g}^1$ frequency difference (~28 cm$^{-1}$) point to the formation of few-layer MoS$_2$ with more than 5 layers [45]. A blue shift of about ~1 and 4 cm$^{-1}$ (Fig. 2(c)) after AuNPs surface decoration were observed for $E_{2g}^1$ and $A_{1g}$ Raman modes of MoS$_2$, respectively. The last behavior is an indicative of involvement of charge transfer processes between MoS$_2$-rGO sheets and dispersed AuNPs, more specifically a p-type doping, where electrons are transferred from MoS$_2$-rGO to AuNPs [46]. Other three peaks at 150, 218 and 328 cm$^{-1}$, attributed to the $J_1$, $J_2$ and $J_3$ longitudinal acoustic modes of S-Mo-S bonds in 1T'-MoS$_2$ phase, are characterizing the Raman spectrum of the MoS$_2$-rGO nanocomposite (Figure 2(b)). Demonstrating the coexistence of 1T' and 2H phases in this sample [7,47]. The coexistence of this two phase in MoS$_2$-rGO nanocomposite was also reported by Jeffery et al., showing that 1T' phase stability is associated to rGO content [48]. After AuNPs surface decoration the intensity of $J_1$, $J_2$ and $J_3$ peaks of 1T' phase becomes much weaker, while the $A_{1g}$ peak (typical of 2H-MoS$_2$ phase) becomes more intense. This result is an indication of phase transformation from metallic 1T' to the semiconducting 2H-MoS$_2$ phase with the presence of AuNPs. The metallic to semiconductor transformation in MoS$_2$-rGO nanocomposite after AuNPs surface decoration could be a promising approach to keep the light absorbing properties of MoS$_2$ with a good electrical conductivity, mechanical stability and high surface area due to the presence of rGO support, making this material an excellent candidate for optoelectronic devices and photocatalytic applications.

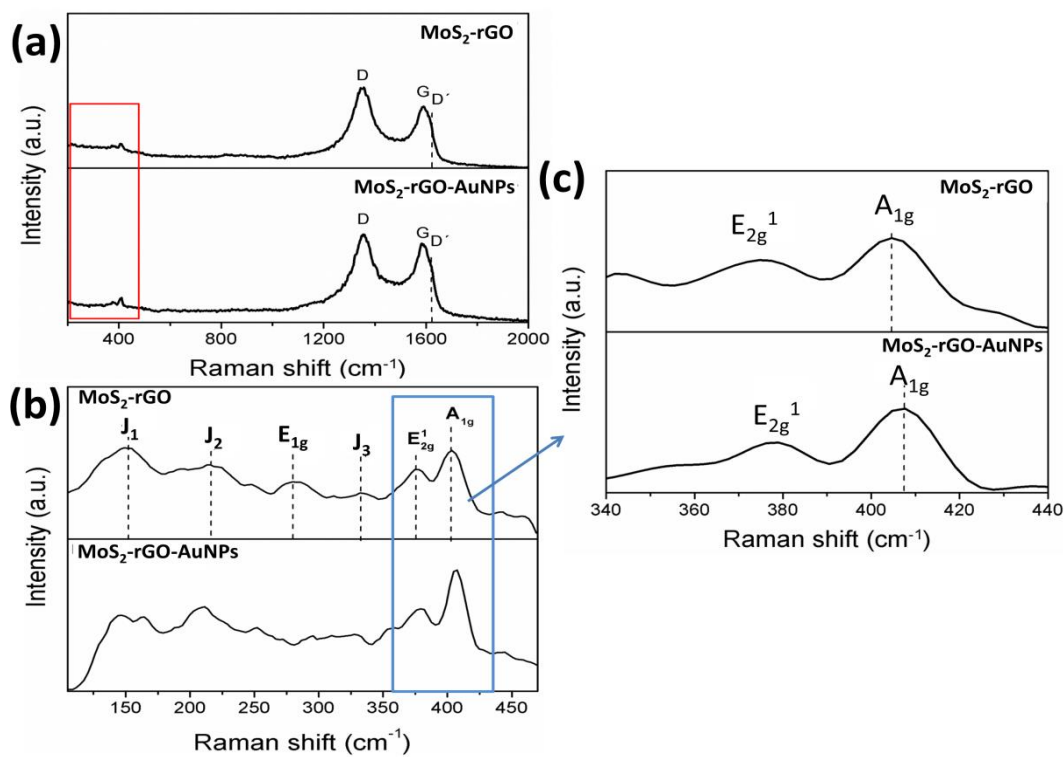

**Figure 2.** (a) Raman spectra of the MoS$_2$-rGO and MoS$_2$-rGO- AuNPs nanocomposites. (b) Raman spectra showing the MoS$_2$ peak regions (inside of red rectangle in figure (a)). (c) Raman spectra showing E$_{2g}^1$ and A$_{1g}$ vibrational modes (inside of blue rectangle in figure (b)).

The chemical composition of the MoS$_2$-rGO and MoS$_2$-rGO-AuNPs nanocomposites was estimated from survey XPS spectra as shown in Figure 3. The core level Mo 3d, S2p, C1s and O1s peaks found in the spectrum of MoS$_2$-rGO confirm the presence of the constituent elements in this nanocomposite. Additionally, Au 4f core level doublet peaks (Au 4f$_{7/2}$ at 84.1 eV and Au 4f$_{5/2}$ at 87.8 eV shown in Figure S3) are observed for the MoS$_2$-rGO-AuNPs sample, confirming the presence of gold nanoparticles in metallic state decorating the MoS$_2$-rGO surface. The atomic percentages for each nanocomposite are reported on Table 1. The S/Mo atomic ratio of ~2, close to the theoretical stoichiometry formula of MoS$_2$ is found for pristine MoS$_2$-rGO. However, a significant decrease (from 17.1% in pristine MoS$_2$-rGO to 8.3% in MoS$_2$-rGO-AuNPs) in the S content is observed after AuNPs surface decoration. The last result may be associated to the fact that the AuNPs on the MoS$_2$-rGO surface are adsorbed on top of the S atoms sites. This result is in agreement with previous theoretical reports where was found that Au adatom favors to adsorb at the top of S atom and with a strong hybridization between the *s* orbital of Au and the Mo 4d states [49,50]. A slight increase of C and O species content is also observed in the MoS$_2$-rGO-AuNPs sample compared with the pristine case, which can be associated to the presence of citrate ligands utilized to stabilize the AuNPs in solution [32].

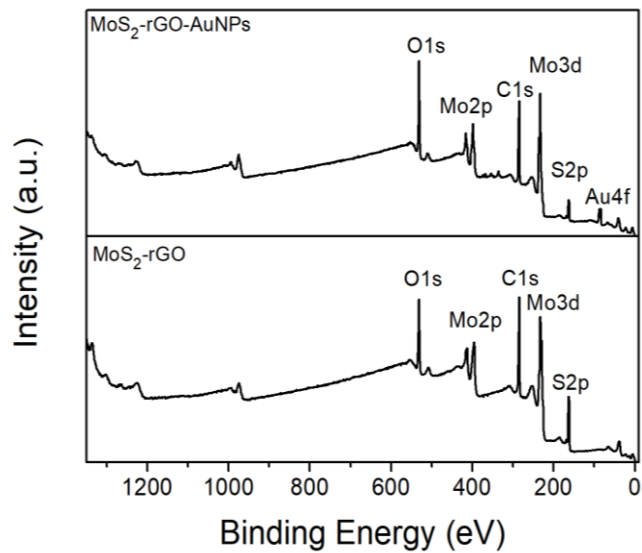

**Figure 3**. XPS survey spectra of MoS$_2$-rGO and MoS$_2$-rGO-AuNPs nanocomposites.

**Table 1**. Surface chemical composition (at.%) distribution of MoS$_2$-rGO and MoS$_2$-rGO-AuNPs nanocomposite films obtained from XPS survey spectra analysis.

|  | C1s | O1s | Mo3d | S2p | Au4f |
|---|---|---|---|---|---|
| MoS$_2$-rGO | 55.4 | 19.6 | 7.8 | 17.1 | - |
| MoS$_2$-rGO-AuNPs | 57.5 | 24.4 | 9.0 | 8.3 | 0.73 |

AuNPs doping effects in the stability of MoS$_2$ (1T' and 2H) phases and in the local electronic structure for the MoS$_2$-rGO nanocomposite were investigated with high resolution XPS spectra analysis. Figure 4 shows the C1s, S2p, Mo 3d and O1s high resolution XPS spectra of both nanocomposites. The fitting parameters, (Binding Energy - BE), full width at half maximum (FWHM) and species amount (% area) of high resolution spectra are summarized in the Table S1. The high resolution C1s XPS spectra of both nanocomposites showed in Figure 4(a) were deconvoluted using three components: peak at BE=284.5 eV can be attributed to C-C/C=C $sp^2$ carbon species, the peak at BE=285.9 eV can be attributed to hydroxyl carbon C-O, and the peak at BE=288.2 eV carboxylate carbon O–C=O [25]. These spectra are similar to the C1s XPS spectrum of isolated rGO reported in the literature [25,28]. There are no significant differences between the C1s XPS spectrum of $MoS_2$-rGO before and after the decoration with AuNPs.

Three sulphur species with their respective spin-orbital doublets are observed in the S2p XPS spectra of both nanocomposites (Figure 4(b)). The peaks at BE = 161.8 eV and 163.1 eV correspond to the S $2p_{3/2}$ and $2p_{1/2}$ sulfur species in $MoS_2$ [51-53]. The second S species observed in the S2p region was attributed to bridging C-S-C structures (BE = 163.3 eV and 164.4 eV to the S $2p_{3/2}$ and $2p_{1/2}$, respectively) [35]. The high energy component at BE = 168.5 eV (S $2p_{3/2}$) can be associated to $S^{6+}$ species in C-$SO_X$-C sulfate groups [52,53].

More representative differences between the electronic structure of $MoS_2$-rGO and $MoS_2$-rGO-AuNPs nanocomposites are observed in the high resolution Mo 3d and O1s XPS spectra. According to previous reports the diverse electronic properties of TMDs arise from the progressive filling of the non-bonding d bands [54,55]. The effect of chalcogen atoms on the electronic structure is minor compared with that of the metal atoms, since the top of the valence band has higher contribution from the Mo orbitals. The phase stability of TMDs also depends of the effective change in the d-electron count in the transition metal atom.

The Mo 3d XPS spectrum of $MoS_2$-rGO nanocomposite is characterized by four contributions associated to different oxidization states and structural phases of the Mo species in this nanocomposite (Figure 4(c)). The peaks at BE = 230.0 eV and 233.2 eV correspond to $Mo^{4+}$ $3d_{5/2}$ and $Mo^{4+}$ $3d_{3/2}$ components of $2H-MoS_2$ (2H) phase of $MoS_2$. The deconvolution of the Mo 3d spectrum identified additional peaks ($Mo^{4+}$ $3d_{5/2}$ at 229.0 eV and $Mo^{4+}$ $3d_{3/2}$ 232.1 eV) appearing at BE ~1 eV below of the corresponding peaks of the 2H phase which were attributed to $1T'-MoS_2$ phase [10,51,52,56-60]. The peaks at 231.2 and 234.4 eV, with separation energy of 3.2 eV, can be attributed to the Mo $3d_{5/2}$ and Mo $3d_{3/2}$ due to the presence of $Mo^{5+}$ [43,53]. In the high energy region of the spectrum, the Mo $3d_{5/2}$ (232.7 eV) and Mo $3d_{3/2}$ (235.8 eV) peaks associated to $Mo^{6+}$ can be found [52, 53]. The peak at BE = 226.3 eV in the low energy region of the Mo 3d spectrum corresponds to S2s contribution. The Mo3d HRXPS results are in accordance with the Raman discussion, showing that $MoS_2$-rGO nanocomposite is composed of a mix of 1T' and 2H $MoS_2$ phases. The same Mo chemical species are observed for the $MoS_2$-rGO-AuNPs sample; however, some important differences are seen concerning their content after AuNPs surface decoration. We notice an increase of $Mo^{6+}$ oxidation species, while the $Mo^{4+}$ (1T')/ $Mo^{4+}$(2H) ratio decreases from 1.7 to 0.6 after the decoration of the $MoS_2$-rGO surface with Au nanoparticles. The increase of $Mo^{6+}$ and the decrease of $Mo^{4+}$ ($1T'-MoS_2$) can be interpreted as electron transfer from $MoS_2$-rGO to Au NPs, converting the $MoS_2$-rGO electronic structure from a metallic to a semiconductor character, corroborating with our Raman spectroscopy results. The increment of $Mo^{6+}$ in $MoS_2$ after AuNPs decoration was also reported by Kim et al. and was attributed to oxidation of $MoS_2$ to molybdic acid [61].

The high resolution O1s spectrum (Figure 4(d) bottom) was deconvoluted by three peaks at BE= 530.6, 531.7 and 533.1eV, respectively. These contributions are associated with C-C=O, C=O/-COOH and OH groups, respectively. After decorating the MoS$_2$-rGO surface with AuNPs, the OH species disappear, which may be attributed to a reduction of the surface adsorbed oxygen (water).

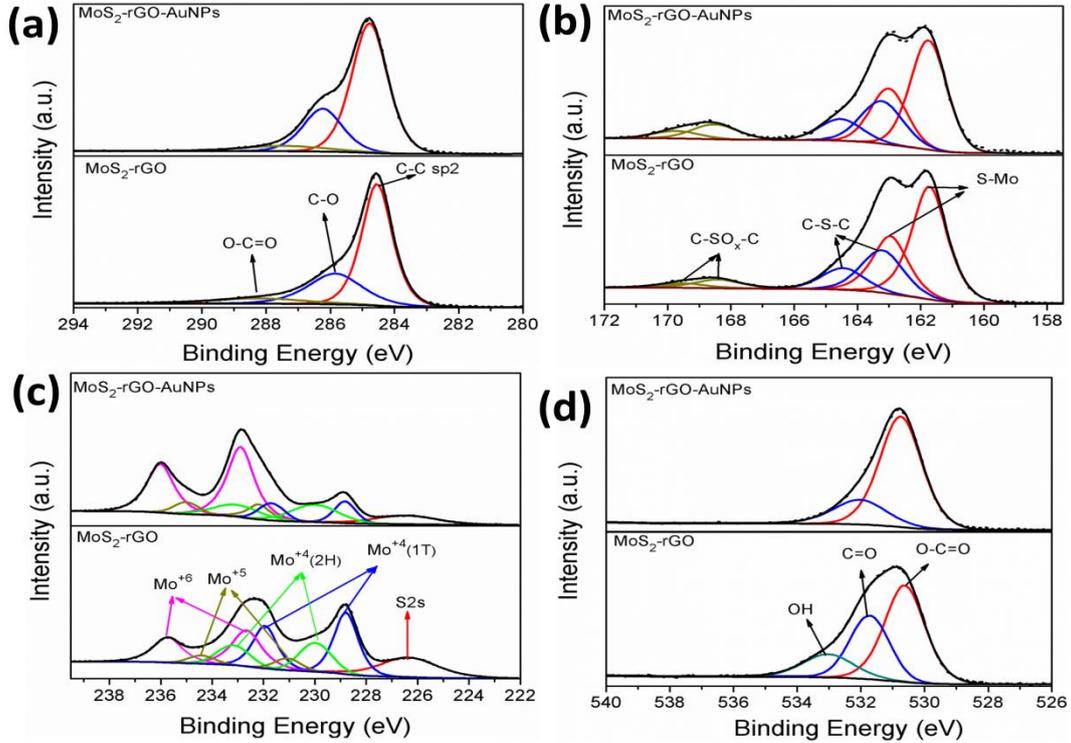

**Figure 4**. High resolution XPS spectra of MoS$_2$-rGO and MoS$_2$-rGO-AuNPs nanocomposites: (a) C1s, (b) S2p, (c) Mo 3d and (d) O1s core levels.

To elucidate the mechanism that induces the phase transition in the nanocomposite, DFT calculations were performed on the different MoS$_2$ phases with or without the presence of gold. It has been proposed that a phase transition can be induced by charge transfer to the TMD [58]. According to Gao, et al., electron injection can stabilize the 1T'-MoS$_2$ phase [58]. However, our experimental results show that electrons are transferred from the MoS$_2$ to the AuNPs reducing the stability of the 1T'-phase and enabling the transition to the more stable 2H-MoS$_2$. We looked at the effects of the interaction between Au and MoS$_2$ in the electronic structure (Figure 5(b)). For the 2H phase adsorbed on an Au (111) surface (see Figure S4) we note little or no effect on the band structure of $MoS_2$ compared to the isolated case. On the other hand, for the 1T'-phase - which is metallic - there is a strong interaction between the metal and the monolayer of $MoS_2$, particularly around the G and M points of the spectrum where the band character becomes intertwined between gold and the TMD indicating a stronger interaction which favors the electrons transfers between the AuNPs and $MoS_2$.

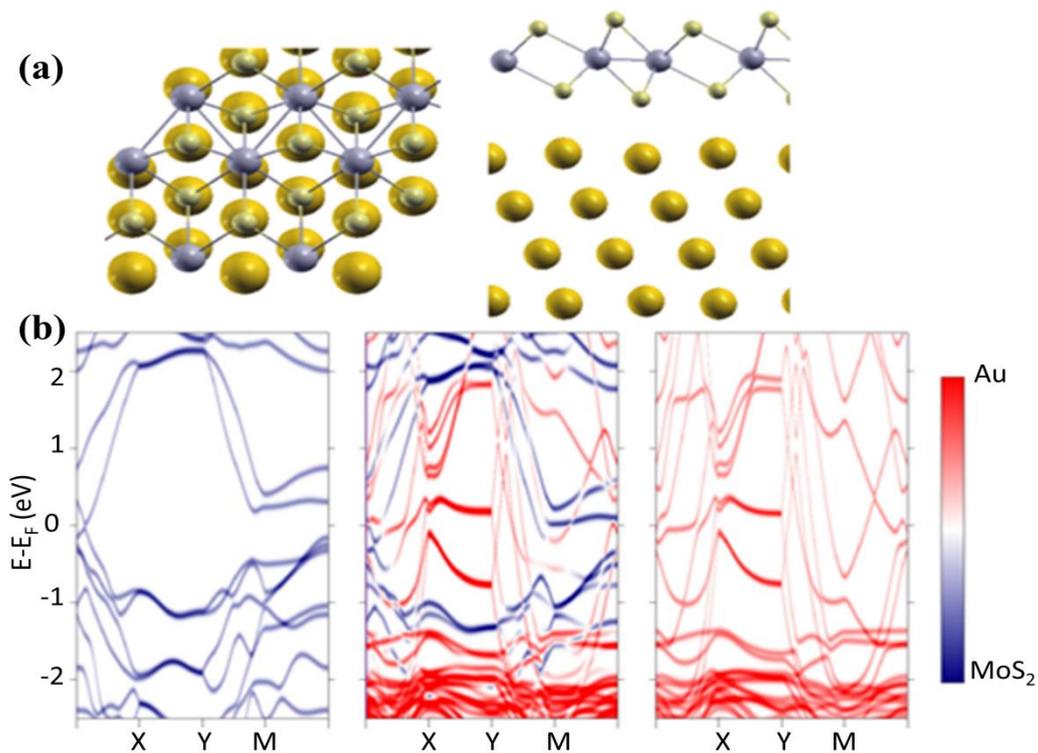

**Figure 5**. (a) Top and lateral view of the 1T' phase of $MoS_2$ adsorbed on a gold (111) surface. (b) Band structure projected on orbitals of the different constituents for the isolated (1T'-$MoS_2$ and Au (111)) structures (outer panels), and the composite material (inner panel).

The valence band (VB) XPS spectra of $MoS_2$-rGO and $MoS_2$-rGO-AuNPs nanocomposites are shown in Figure 6. The valence band spectrum for $MoS_2$-rGO is qualitatively similar to $MoS_2$ stemming from the hybridization of Mo4d and S3p electronic states [61]. According to crystal-field theory, Mo 4d orbitals are split into three orbitals (see the Figure 6(c)): one with low energy $4d_{z^2}$4d, two doubly degenerated composed by $4d_{x^2-y^2,xy}$4d(empty) and two doubly degenerated $4d_{xz,yz}$ (empty) for 2H-$MoS_2$ phase (in hexagonal $D_{3h}$ configuration). In this case Mo $4d_{z^2}$ is fully occupied by two spin paired electrons forming the VB. However, 1T'-$MoS_2$ (octahedral $O_h$ (1T')) is composed by three degenerated orbitals $4d_{xy, xz, yz}$ and doubly degenerated $4d_{x^2-y^2,z^2}$ VB in 1T'-$MoS_2$ is composed by partially occupied $4d_{xy,xz,yz}$ [61, 62]. The main effect of AuNPs surface decoration is observed through the decreasing of the peak at ~ 2 eV, close to valence band edge. This behavior can be associated to an electron transfer from Mo 4d states to AuNPs conduction band how was depicted from Mo3d HRXPS and Raman spectra analyzed previously. Therefore, after a careful analysis of the valence band region and from theoretical calculations by DFT, comparing the electronic coupling between Au and $MoS_2$ states for 1T' and 2H phases (Figure 5(b) and Figure S4) we can conclude that a possible mechanism for the phase transformation from 1T' to 2H-$MoS_2$ that arises after surface decoration is the output of electrons specifically from the Mo $4d_{xy,xz,yz}$ that make up the VB in phase 1T'.

Figure 6(b) shows the extrapolation procedure to determine the valence band maximum edge for both nanocomposites from valence band XPS spectra. A shift ~ 0.1 eV to lower BE values is observed after nanoparticle decoration. This shift is an indicative of p-type doping, where the AuNPs behave as electron acceptors from $MoS_2$-rGO valence band [61,62]. This result is in agreement with Mo3d XPS analysis where the electron transfer from $MoS_2$-rGO to AuNPs decreased their metallic character.

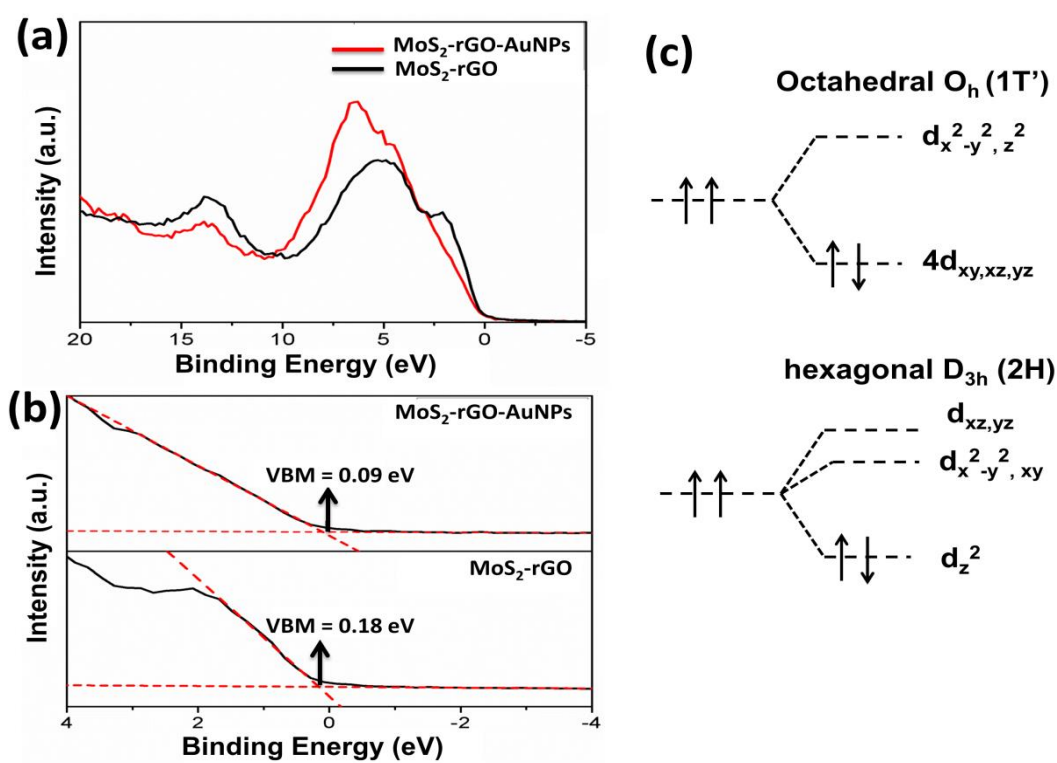

**Figure 6**. (a) Valence band XPS spectra of $MoS_2$-rGO and $MoS_2$-rGO-AuNPs nanocomposites. (b) Extrapolation procedure to determine the valence band maximum level. The inset show the electronic configuration induced by crystal-field splitting of octahedral $O_h(1T')$ and $D_{3h}(2H)$ $MoS_2$ phases.

The partially unoccupied states were investigated with S $K$-edge NEXAFS spectroscopy. Figure 7 shows the S $K$-edge NEXAFS spectra of $MoS_2$-rGO and $MoS_2$-rGO-AuNPs nanocomposites. The $MoS_2$-rGO NEXAFS spectrum shows two main peaks below the sulfur ionization energy. Peak 2 (2471 eV) at resonance maximum can be attributed, according to previous reports, [63-65] to electronic transitions from the S1s core level to unoccupied hybridized S3p-Mo4d electronic states. The pre-edge shoulder at 2468.5 eV (peak 1) can be associated with the presence of S-O bond species [52,60,66]. A third peak, located at 2472.1 eV and attributed to S1s–$3p_z$ electronic transitions has been observed for the $MoS_2$ monolayer sample (black line spectrum in Figure 7) showed here as comparison [63,65]. However, peak 3 is poorly defined in the NEXAFS spectra of $MoS_2$-rGO and $MoS_2$-rGO-AuNPs nanocomposites. According to

Guay et al., the energy separation between S3p$_{x,y}$ and S3p$_z$ electronic states is due to the difference of their symmetric properties at symmetry points Γ and A of the Brillouin zone for the 2H-MoS$_2$ phase [65]. Considering the Mo3d XPS results we saw that MoS$_2$-rGO nanocomposites are a mixture between 2H-MoS$_2$ and 1T'-MoS$_2$ phases, then, the presence of 1T'-MoS$_2$ metallic phase changes the symmetry properties of S2p electronic states and reduces the energy separation between S3p$_{x,y}$ and S3p$_z$.

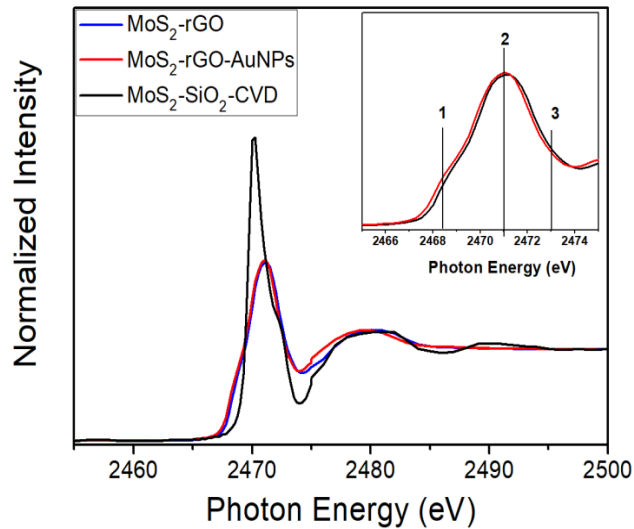

**Figure 7.** S *K*-edge NEXAFS spectra of MoS$_2$-rGO (blue solid line) and MoS$_2$-rGO-AuNPs (red solid line) nanocomposites. The S *K*-edge NEXAFS spectrum of the MoS$_2$ monolayer obtained by CVD [63] (black line) is also shown for the purpose of comparison. The inset shows a zoom of the first resonance peak. The numbers indicate the photon energies used to obtain the S-K L$_{2,3}$L$_{2,3}$ RAS spectra.

The degree of delocalization or electronic coupling between unoccupied electronic states is investigated by the core hole clock approach using S-K L$_{2,3}$L$_{2,3}$ RAS spectra. A detailed explanation about core hole clock method can be found elsewhere [67, 68] and is summarized in the supporting information file (SI). In this approach, the charge transfer times, used as a quantitative measure of electronic coupling (lower charge transfer time means high electronic coupling), are determined using lifetime

($\tau_{CH}$=1.27 fs in the case of the S *K*-edge [69]) of the core electron hole as an internal reference clock by:

$$\tau_{CT}=(I_{Raman}/I_{CT-Auger})\times \tau_{CH},$$

where $I_{Raman}$ and $I_{CT-Auger}$ represent the intensities of different core-hole decay channels achieved by S-K $L_{2,3}L_{2,3}$ RAS spectroscopy. The "Raman" and "CT-Auger" electron decay channels correspond to different electronic final states. Raman decay correspond to the spectator two hole and one electron final states (2h1e) where the excited electron does not participate in the decay process. The second CT-Auger decay process consists of two holes (2h) in the valence band, reached when the electron is transferred out of the atom during the core-hole lifetime. S-K $L_{2,3}L_{2,3}$ RAS spectra shown in Figure 8 are convoluted by these core-decay channels. The attribution of "Raman" and "CT-Auger" decay channels in the RAS spectra is based on their behavior when the incident photon energy is tuned across the core excitation resonance. In Raman decay channels, the kinetic energy of the emitted electron increases when the energy of an incident photon is tuned across the resonance, while the CT-Auger contribution keeps at constant kinetic energy independently of the energy of the incident photon. The identification of the Raman and CT-Auger contributions forming the S-K $L_{2,3}L_{2,3}$ RAS spectra for MoS$_2$ was reported recently by Garcia-Basabe et al [63]. According to the last mentioned behavior of the "Raman" and "CT-Auger" decay channels (showed in Figure S5), we identified two spectators (blue and green in Figure 8) and one CT-Auger (red) contributions.

In this way, we determined $\tau_{CT}$ for S-K $L_{2,3}L_{2,3}$ RAS spectra obtained at photon energies labelled as 1 (2468.5 eV), 2 (2471 eV) and 3 (2472.1 eV) in the inset of Figure 7. The excitation of an electron to an unoccupied electronic state in the pre-edge shoulder of the S1s resonance spectrum (labelled as 1 (hυ=2468.5 eV)), the $\tau_{CT}$ of 5.38 ± 0.60 fs and 7.74 ± 0.70 fs were obtained for MoS$_2$-rGO and MoS$_2$-rGO-AuNPs nanocomposites, respectively. The increase in the $\tau_{CT}$ value for this photon energy was due to a decrease of the CT-Auger contribution from 18% in $MoS_2$-rGO to 12% in $MoS_2$-rGO-AuNPs nanocomposite. On the other hand, for electrons excited to electronic states in the S $K$-edge resonance maximum (labelled as 2 (hυ=2471.0 eV)) the contribution of CT-Auger decay channel decrease from 24% to 17%, while the SP1 decay channel increases from 56% to 64% after AuNPs decoration. Therefore, for excitation energy of 2471.0 eV the $\tau_{CT}$ increase from 3.67 ± 0.60 fs in the $MoS_2$-rGO to 5.52 ± 0.70 fs in the $MoS_2$-rGO-AuNPs nanocomposites. The increase of $\tau_{CT}$ observed in the $MoS_2$-rGO-AuNPs nanocomposite can be interpreted as a reduction of the metallic character of the $MoS_2$-rGO nanocomposite due to the AuNPs. However, for photon energies labelled as 3 (hυ=2472.1 eV), the $\tau_{CT}$ are very similar for both nanocomposites, close to 2.90 fs. The last results mean that the AuNPs affects only the unoccupied electronic states near to the $MoS_2$-rGO Fermi level. Thus, the results obtained by the core hole clock approach are in agreement with Mo 3d XPs analysis, where a transformation from metallic to semiconductor character after AuNPs decoration was observed.

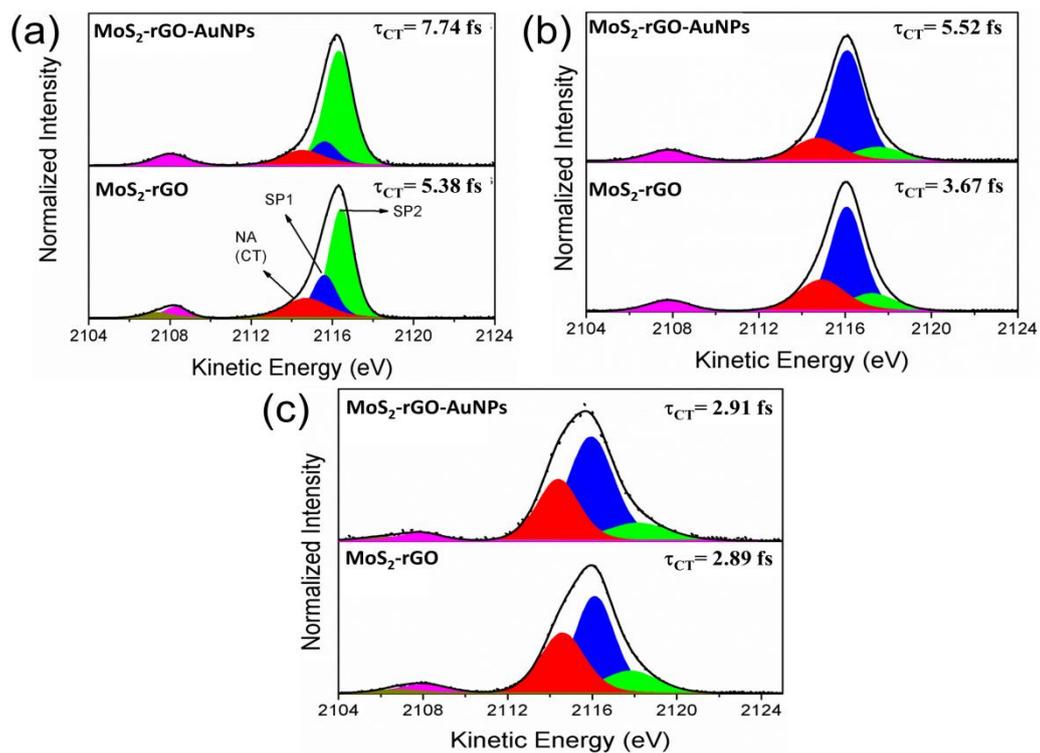

**Figure 8**. S-*K* $L_{2,3}L_{2,3}$ RAS spectra measured at photon energies labelled as 1-3 (2468.5 (a), 2471(b) and 2472.1 eV (c), respectively) in S *K*-edge NEXAFS spectrum. RAS spectra deconvolution in Raman spectators (blue - SP1 and green - SP2) and CT (red) decay channel are also showed.

## 4. CONCLUSION

In summary, we propose a simple approach to transform the $MoS_2$ from metallic to semiconductor character by AuNPs surface decoration in the solvothermal synthesized $MoS_2$-rGO nanocomposite. The AuNPs induces a p-type doping in $MoS_2$-rGO nanocomposite and reduce the stability of 1T' phase favoring the transformation to 2H-$MoS_2$. The main mechanism for this phase transition was electron transfer from 1T' Mo $4d_{xy,xz,yz}$ electronic states to AuNPs conduction band. The unoccupied electronic structure is also affected by AuNPs decoration, showing a strong electronic coupling between AuNPs and $MoS_2$ states close to the Fermi level. Our results suggest a practical method to modulate metallic and semiconductor character of $MoS_2$ phases in $MoS_2$-rGO nanocomposite.

## ASSOCIATED CONTENT

**Supporting Information**

The SEM images and EDX spectra collected at different points in the $MoS_2$-rGO and $MoS_2$-rGO-AuNPs nanocomposites, Au 4f HRXPS spectrum of $MoS_2$-rGO-AuNPs nanocomposite, fitting parameters obtained from C1s, O1s, S2p and Mo3d high resolution spectra deconvolution for the $MoS_2$-rGO and $MoS_2$-rGO-AuNPs nanocomposites, band structure projected on orbitals of the different constituents for the isolated (2H-$MoS_2$ and Au (111)) structures and the composite material, Core Hole Clock Approach description section and photon energy dependence of electron kinetic energy of Resonant Auger decay channels.

## ACKNOWLEDGEMENTS

Y. Garcia-Basabe acknowledges to CNPq for financial support (project (nº408265/2016-7). Research partially supported by LNLS – National Synchrotron Light Laboratory (SXS-20160596) and by CNPEM-LNNano (XPS-21999), Brazil. A.R.R. acknowledges financial support from CNPq and FAPESP (Project nº 2012/50259-8) and the Simons Foundation through the ICTP Associateship program. Theoretical calculations were carried out in the Scientific Computing Center at UNESP (NCC-Unesp), and in the Santos Dumont Supercomputer (LNCC-Santos Dumont). D. Grasseschi acknowledges to FAPESP for the financial support (grant no. 2015/10405-3).